\documentclass[10pt,conference]{IEEEtran}
\IEEEoverridecommandlockouts
\IEEEoverridecommandlockouts
\usepackage{cite}
\usepackage{amsmath,amssymb,amsfonts}
\usepackage{algorithm}
\usepackage{algpseudocode}
\usepackage{graphicx}
\usepackage{textcomp}
\usepackage{booktabs}
\usepackage{listings} % For code formatting
\usepackage{xcolor} % For syntax highlighting

\def\BibTeX{{\rm B\kern-.05em{\sc i\kern-.025em b}\kern-.08em
    T\kern-.1667em\lower.7ex\hbox{E}\kern-.125emX}}

\begin{document}

\title{OSPtrack: A Labeled Dataset Targeting Simulated Execution of Open-Source Software}

\author{\IEEEauthorblockN{Zhuoran Tan}
\IEEEauthorblockA{\textit{University of Glasgow} \\
UK \\
z.tan.1@research.gla.ac.uk}
\and
\IEEEauthorblockN{Christos Anagnostopoulos}
\IEEEauthorblockA{\textit{University of Glasgow} \\
UK \\
Christos.Anagnostopoulos@glasgow.ac.uk}
\and
\IEEEauthorblockN{Jeremy Singer}
\IEEEauthorblockA{\textit{University of Glasgow} \\
UK \\
jeremy.singer@glasgow.ac.uk} 

}

% \author[$\dag$]{Taylor Brierley\thanks{Email: taylor.brierley@jumpsec.com}}
% \author[*]Sebastian Garcia\thanks{Email: sebastian.garcia@agents.fel.cvut.cz}}


\maketitle

\begin{abstract}
Open-source software serves as a foundation for the internet and the cyber supply chain, but its exploitation is becoming increasingly prevalent. While advances in vulnerability detection for OSS have been significant, prior research has largely focused on static code analysis, often neglecting runtime indicators. To address this shortfall, we created a comprehensive dataset spanning five ecosystems, capturing features generated during the execution of packages and libraries in isolated environments. The dataset includes 9,461 package reports, of which 1,962 are identified as malicious, and encompasses both static and dynamic features such as files, sockets, commands, and DNS records. Each report is labeled with verified information and detailed sub-labels for attack types, facilitating the identification of malicious indicators when source code is unavailable. This dataset supports runtime detection, enhances detection model training, and enables efficient comparative analysis across ecosystems, contributing to the strengthening of supply chain security.

\end{abstract}


\section{Introduction}

\IEEEPARstart{O}{pen} source software (OSS) has become a foundational element in a wide range of applications, but its integration has introduced vulnerabilities increasingly exploited in adversary activities \cite{mandianttrends2024}. Existing vulnerability detection methods often focus on static code analysis \cite{ladisaDetectionMaliciousJava2022,10.1145/3691621.3694950}, which can be inadequate when OSS is embedded within complex systems. Current datasets typically focus on malicious packages \cite{OpenSourceDatasetMaliciousSoftwarePackages,10.1007/978-3-030-52683-2_2,trandetectvul2024}, but fail to capture runtime features. Although software simulations have been developed, no dataset specifically targets OSS and provides labeled monitoring results during execution.

This dataset leverages the package-analysis tool \footnote{https://github.com/ossf/package-analysis} to simulate package execution, using authenticated references from existing datasets in the supply chain security domain \cite{malicious-packages,10.1007/978-3-030-52683-2_2}. By parsing the generated reports and extracting features, we provide a comprehensive labeled dataset with eight-dimensional features, including files, sockets, commands, and DNS-related behaviors.

The dataset, generated over five weeks, includes 9,461 valid items across ecosystems such as npm, pypi, crates.io \footnote{https://crates.io/}, nuget \footnote{https://www.nuget.org/}, and packagist \footnote{https://packagist.org/}. The data, along with the code to simulate package execution, feature extraction, and label matching, are available online \cite{tan_2024_14197378} to facilitate reproducibility and further research.

\section{Background}

\subsection{OSS Supply Chain Security}

The OSS has become a critical component of the supply chain. Since SolarWinds attack \cite{9382367},  the exploitation of OSS vulnerabilities has grown increasingly prevalent, prompting the development of numerous mitigation strategies.

Ladisa et al. \cite{ladisaDetectionMaliciousJava2022} identified the means of spreading malware through the OSS supply chain. By analyzing dataset in \cite{ohmBackstabberKnifeCollection2020}, they uncovered several exploitation cases, including reverse shell spawning a shell process, dropper with second-stage payload, and data exfiltration, among others.

Zhang et al. \cite{zhang_malicious_2023} present a categorization of detection methods for malicious packages. They identify three main categories: rule-based, unsupervised, and supervised learning approaches. Rule-based methods prioritize precise matching of IOCs associated with malicious behaviors, such as specific IP addresses, domain names, or registry values. This method is widely used in industry because of its accuracy.

Liang et al. \cite{10.1145/3691621.3694950} explored the detection for mobile applications. The proposed method created graph representation of code property graphs (CPGs) from source code and the detection combines advanced machine learning algorithms like CodeBERT, and attention of Transformers with Graph Convolutional Networks (GCN).

Tran et al. \cite{trandetectvul2024} introduced a BERT-based architecture to detect vulnerable statements in Python code. This framework is capable of learning complex relationships between code statements without predefined graph structures.

Most existing research on OSS vulnerabilities\cite{guo2023empiricalstudymaliciouscode,rozi_securing_2024,tran_detectvul_2024} primarily focuses on analyzing plain code, functions, or dependencies. However, when it comes to runtime behaviors, related studies tend to concentrate on detecting malware or ransomware\cite{Parisot2024RansomwareDL,juan_a__herrera_silva__2023}, rather than addressing the detection of malicious scripts embedded within OSS.

\subsection{Threat Data Generation}

Zhang et al. \cite{zhang_malicious_2023} highlighted learning-based approaches for extracting features via simulation, dynamic, or static analysis, emphasizing three key scenarios for triggering malicious behavior: install-time, import-time, and run-time. These insights guide our use of simulation to capture such behaviors.

Landauer et al. \cite{landauer_have_2021} simulated the full cyber kill chain, including reconnaissance via scanning, delivery through webshell exploitation, and achieving root privilege for command execution. While this comprehensive approach captured diverse exploitation behaviors, its design complexity limited the number of scenarios, reducing generality.

Parisot et al. \cite{Parisot2024RansomwareDL} introduced a novel approach for ransomware detection by combining the dynamic analysis features of the Cuckoo Sandbox \footnote{https://cuckoosandbox.org/} with machine learning techniques. Their method involved creating three datasets that capture API calls, network interactions, and string data from malware samples, with feature extraction using the TF-IDF technique \cite{aizawa2003information}.

Silva et al. \cite{juan_a__herrera_silva__2023} employed dynamic analysis and machine learning to identify ransomware signatures using selected dynamic features. The research involves experiments with encryptor and locker ransomware, alongside goodware, to generate a dataset of dynamic parameters in a controlled environment using a sandbox. 

Several datasets have been released for OSS research and defense purposes\cite{OpenSourceDatasetMaliciousSoftwarePackages,10.1007/978-3-030-52683-2_2,malicious-packages}. However, these datasets primarily focus on collecting malicious packages and libraries, rather than analyzing their runtime behaviors. While a public dataset has been generated using the tool package-analysis, it lacks detailed labeling and specific feature information, which significantly limits its utility.

\section{Methodology}

\begin{figure*}[ht]
    \centering
    \includegraphics[scale=0.73]{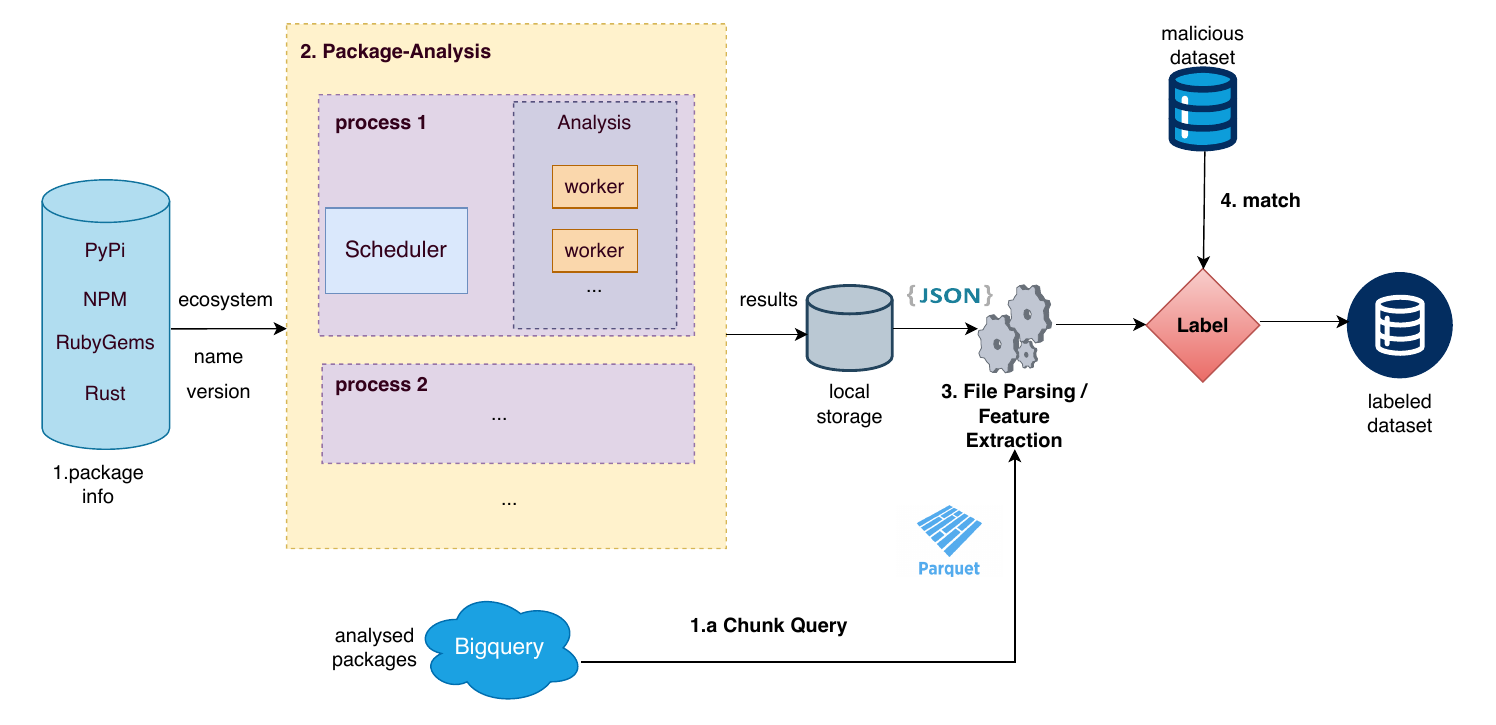}
    \caption{The Data Generation Framework. 1. Collect package information 1.a. Query analyzed results from BigQuery 2. Simulate packages with package-analysis in multiple processes 3. Parse JSON reports and also queried Parquet reports, extract features 4. Match and generate labels based on known labels }
    \label{fig: framework}
    \end{figure*}

To generate a labeled OSS-tailored dataset focusing on running-time features, we adopt the methodology demonstrated in Fig. \ref{fig: framework}. The core design includes parallelized  analysis for packages, report parsing with feature extraction, and final label matching. Additionally, part of dataset, especially majority of benign examples, is from the queried results from available BigQuery dataset generated by package-analysis tool. The detailed pipelines are illustrated below.

\subsection{Package Collection}

Two key datasets are included when collecting malicious package information: the Backstabbers-Knife-Collection (BKC) \cite{ohmBackstabberKnifeCollection2020} and the malicious packages repository \cite{malicious-packages}. The BKC dataset, comprising 5,942 samples from npm, PyPI, and Ruby, is widely cited in academic studies \cite{guo2023empiricalstudymaliciouscode, ladisaDetectionMaliciousJava2022}. The second dataset, created by the Open Source Security Foundation, is the most comprehensive OSS dataset, covering the majority of documented supply chain attacks up to November 2024. It includes 24,215 samples across additional ecosystems such as crates.io, NuGet, and Packagist.

The malicious packages in both datasets are organized using JSON files that provide package details, including the name, version, ecosystem, indicators of compromise (IOCs), and exploitation methods. Leveraging core indicators such as blockchain activity, source code, and commands described in the JSON files, we further categorize the potential attack type, malicious functions, trigger mechanisms, code locations, and evasion techniques, following the methodologies outlined in \cite{ohmBackstabberKnifeCollection2020,zhang_malicious_2023}.

After aggregating all the data, two files are generated: \texttt{data\_iocs.csv} and \texttt{data\_metrics.csv}. The IOC file compiles indicators such as domains, IP addresses, links, URLs, and strings, serving as a dataset for directly identifying malicious packages or libraries. The metrics file consolidates detailed information on malicious packages, enabling the exploration of high-level package behaviors and trends in package exploitation across diverse ecosystems. Additionally, the metrics file facilitates label and sub-label matching due to the detailed package specifications it contains. Statistical insights derived from the metrics file are summarized in Table \ref{table:attack types of supply chain attack}.

\begin{table}[ht]
\caption{Collection of Attack Types in Software Supply Chain}
\label{table:attack types of supply chain attack}
\centering
\begin{tabular}{ll}
\hline
\textbf{Attack Type}                      & \textbf{Number} \\ \hline
data exfiltration/root shell     & 12026   \\ 
dropper/malware                  & 5425   \\ 
typosquatting                    & 937    \\ 
command \& control(C2) communication                 & 1674    \\ 
malicious command execution      & 118     \\ 
starjacking                      & 31     \\ 
social engineering/typosquatting & 14     \\ 
typosquatting/malware & 1     \\ \hline

\end{tabular}
\end{table}

A portion of the dataset is sourced from the BigQuery dataset generated by the package-analysis tool. This public dataset is based on live instances of newly published packages or libraries and include static and dynamic features. However, the dataset does not include explicit labels for the reports, so labeling still needs to be confirmed.
% The code we query data from BigQuery has been provided as follows:

% \begin{lstlisting}
%     SELECT *
%     FROM `ossf-malware-analysis.packages.analysis`
%     LIMIT {chunk_size}
%     OFFSET {order}
% \end{lstlisting}

In total, we collect 7,500 package reports from this dataset, which constitutes the majority of benign samples. These reports are saved in Parquet \footnote{https://parquet.apache.org/} format to preserve the original data types, such as lists or dictionaries, rather than converting them to strings as in CSV format. This approach eliminates the need for additional transformations and complex string-parsing operations. Another part of dataset is generated through calling package-analysis tool to simulate the execution of known malicious packages.

\subsection{Simulation}

Package-analysis is a sandbox tool used to simulate package execution. Sandboxes have been widely adopted to analyze the dynamic behaviors of software and detect malicious activity \cite{10.1145/2976749.2989064,10529261}, with notable examples including Cuckoo and VirusTotal \cite{shuofei_zhu__2020}. Unlike these sandboxes, package-analysis allows for inputting package information such as ecosystem, name, and version, and provides the output of real code or packages for analysis.

The analysis results generated by this tool consist of both dynamic and static analysis derived from the simulated execution of malicious packages. The dynamic analysis includes a comprehensive collection of HTTP or HTTPS interactions, combining DNS queries, command executions, socket connections, standard errors (stderr), file manipulations, status updates, and standard output (stdout). In contrast, the static analysis gathers direct information such as suspicious identifiers, escaped strings, base64 strings, hex strings, IP addresses, and URLs.

The scheduler in this tool creates analysis jobs from package feeds, with the analysis component performing one-time tasks while workers collect package behavior data through static and dynamic analysis. By default, the results are pushed to BigQuery for public access. To store results locally, we modified the simulation script to save files in the format {ecosystem}-{package}-{version}. Additionally, to prevent duplicates, we implemented a check to ensure that previously generated results are not saved again.

The simulation runtime varies from several seconds to more than 20 minutes, depending on the size of the analyzed package. Some packages, as recorded in \cite{malicious-packages} and \cite{10.1007/978-3-030-52683-2_2}, were unable to be analyzed due to missing dependencies. Furthermore, certain packages caused the simulation to become stuck, affecting the analysis of subsequent packages in the queue. To address this, we implemented a timeout mechanism (e.g., 10 minutes) and regular health checks to detect and mitigate simulation stalls.

In terms of performance, a single process generates an average of daily 50 package reports. By implementing parallelization, each starting from different row indices, the generation speed increases to approximately daily 150 packages. In five weeks, 4235 reports have been created, in which only 1961 is valid. We can verify the large proportion of insufficient simulations. The main reason behind this is missing, mitigated, or removed historical packages after identifying vulnerabilities. Those reports are in JSON format and are processed in the next stage for parsing and feature extraction.

\subsection{Feature Extraction}

After collecting both reports from package-analysis and BigQuery dataset, the next pipeline is the parsing engine, in which feature extraction is included. 
The feature extraction involves extracting package information with package name, version, and its belonging ecosystem from \textit{Package} key, and the rest detailed features from system and network interactions, which include \textit{execution}, \textit{import}, and \textit{install} behaviors. Under both \textit{import} and \textit{install} sections, the collected features include \textit{Status, Stdout, Stderr, File, Sockets, Commands, and DNS.} However, the values of some keys are encoded random code (\textit{Stdout, Stderr}) or contain non-important information (\textit{Status}). After bypassing those unnecessary features, we choose only the remaining four-dimensional features, distributing in two categories (\textit{import, install}), resulting in eight features out of package-related features. 

\subsection{Labeling Method}

The labeling process has been divided into two separate parts in order to process separate data from simulation and BigQuery. As demonstrated in Algorithm \ref{alg:label_matching}, 
the labeling has also included two layers, in which the first layer defines malicious(1) or benign(0) and the second layer denotes specific attack types under malicious category. The matching references are generated through two available dataset \cite{ohmBackstabberKnifeCollection2020,malicious-packages}, denoted as \texttt{bkc\_mal.csv} and \texttt{pkg\_mal.csv} in dataset. 

\begin{algorithm}[ht]
\caption{Label Matching Algorithm}
\label{alg:label_matching}
\begin{algorithmic}[1]
\Require $D_1$, $D_2$, $R_f$ (Matching Reference), $M_f$ (Metric)
\Ensure Combined Dataset with Updated $L$ (Labels) and $SL$ (Sub-Labels)

\State \textbf{Initialize:} Load $D_1$, $D_2$, $R_f$, and $M_f$.

\ForAll{records in $D_1$}
    \State Extract $(P, E, V)$ from record. \hfill \textit{($P$: Name, $E$: Ecosystem, $V$: Version)}
    \State Search $R_f$ for matching $(P, E, V)$.
    \If{match found}
        \State Update $L$ in $D_1$ using $R_f$.
    \EndIf
\EndFor

\ForAll{records in $D_2$ (default $L = 1$)}
    \State Extract $(P, E, V)$ from record. 
    \State Search $M_f$ for matching $(P, E, V)$.
    \If{match found}
        \State Assign $SL$ using \texttt{attack\_type} in $M_f$.
    \EndIf
\EndFor

\State Combine $D_1$ and $D_2$, preserving all records and updated $L$, $SL$.
\State \textbf{Output:} Combined dataset with updated $L$ and $SL$.
\end{algorithmic}
\end{algorithm}

\section{Dataset Description}

The entire simulation process and experiments are conducted on a virtual machine running Ubuntu 22.04.5 LTS with 8 gigabytes of memory.
The generated labeled dataset includes 7499 benign examples and 1962 malicious examples, spanning five ecosystems. The dataset is stored in pickle \footnote{https://docs.python.org/3/library/pickle.html} format to remain the original data type. The distribution of labels has been illustrated in Table \ref{tab:combined_package_counts}, where it can be observed that the majority of packages are from the npm ecosystem, while Ruby represents the smallest proportion.

\begin{table}[ht]
\centering
\caption{Package counts by ecosystem, package count, label, and sub-label.}
\begin{tabular}{lcclc}
\toprule
\textbf{Ecosystem} & \textbf{Count} & \textbf{Label} & \textbf{Sub\_Label} \\ 
\midrule
crates.io  & 1205 & 0 & na              \\ 
npm        & 4645 & 0 & na              \\ 
packagist  & 265  & 0 & na              \\ 
pypi       & 1323 & 0 & na              \\ 
rubygems   & 61   & 0 & na              \\ 
crates.io  & 1    & 1 & na              \\ 
npm        & 800  & 1 & na              \\ 
pypi       & 812  & 1 & na              \\ 
rubygems   & 269  & 1 & na              \\ 
npm        & 18   & 1 & C2              \\ 
pypi       & 38   & 1 & C2              \\ 
rubygems   & 8    & 1 & C2              \\ 
npm        & 11   & 1 & root shell      \\ 
npm        & 2    & 1 & command exec    \\ 
pypi       & 2    & 1 & command exec    \\ 
\bottomrule
\end{tabular}
\label{tab:combined_package_counts}
\end{table}

Apart from the main labeled dataset, we also include extracted files from available datasets. Those files can be further classified into statistical data and information data. The first group of statistical data denote the iocs file and metrics file created from dataset \cite{malicious-packages}. They are used for matching sub labels and statistic analysis on malicious behaviors. The information data denotes the package information collected from two datasets, playing as references to match malicious labels for queried data.

\section{Research Opportunities}

The dataset is primarily designed for supply chain security research. It can be used to train machine learning models to differentiate between benign and malicious software behaviors, as well as detect vulnerabilities in running software to ensure supply chain security in open-source software (OSS).

Additionally, it is possible to identify indicators triggered by malicious activity, which can aid in detection and prevent further exploitation. The dataset's diversity also enables differential or comparative analysis of malicious package behaviors across different ecosystems \cite{barr-smithExorcistAutomatedDifferential2022}.

Moreover, more detailed feature extraction can be performed on this dataset, allowing for better attack behavior profiling \cite{TANG2022108261} and facilitating representation learning, such as graph-based representations \cite{9834133,LIU2024102748}.

\section{Limitations}

Since the source code is unavailable, the simulated scenarios cannot fully capture the injection process at the source due to the wide variety of exploitation methods. Additionally, some malicious packages could not be analyzed or have reports generated due to missing or deleted packages. Packages requiring extended analysis time were excluded from the dataset because of timeout settings. Furthermore, only a portion of the available malicious packages was simulated due to time constraints. We plan to regularly update the dataset to include a more extensive and diverse collection of malicious reports.

\section{Conclusion}

This work introduces a running-time-oriented labeled dataset for supply chain security research, featuring ground-truth labeling and sub-labels for attack types. It covers five ecosystems—NPM, Packagist, PyPI, RubyGems, and Crates.io—and includes eight extracted features representing both static and dynamic behaviors. The dataset is designed for vulnerability detection and malicious software classification. Future work will focus on expanding the dataset with additional examples for specific ecosystems and exploring its application in vulnerability detection.

\section*{Acknowledgments}
This research has received funding from JUMPSEC Ltd.

% {\appendix[Proof of the Zonklar Equations]
% Use $\backslash${\tt{appendix}} if you have a single appendix:
% Do not use $\backslash${\tt{section}} anymore after $\backslash${\tt{appendix}}, only $\backslash${\tt{section*}}.
% If you have multiple appendixes use $\backslash${\tt{appendices}} then use $\backslash${\tt{section}} to start each appendix.
% You must declare a $\backslash${\tt{section}} before using any $\backslash${\tt{subsection}} or using $\backslash${\tt{label}} ($\backslash${\tt{appendices}} by itself
%  starts a section numbered zero.)}


\clearpage
% \bibliographystyle{IEEEtran}

 

% \begin{IEEEbiography}%[{\includegraphics[width=1in,height=1.25in,clip,keepaspectratio]{fig1}}]{Michael Shell}
% \end{IEEEbiography}



% \appendix


\begin{thebibliography}{10}
\providecommand{\url}[1]{#1}
\csname url@samestyle\endcsname
\providecommand{\newblock}{\relax}
\providecommand{\bibinfo}[2]{#2}
\providecommand{\BIBentrySTDinterwordspacing}{\spaceskip=0pt\relax}
\providecommand{\BIBentryALTinterwordstretchfactor}{4}
\providecommand{\BIBentryALTinterwordspacing}{\spaceskip=\fontdimen2\font plus
\BIBentryALTinterwordstretchfactor\fontdimen3\font minus \fontdimen4\font\relax}
\providecommand{\BIBforeignlanguage}[2]{{%
\expandafter\ifx\csname l@#1\endcsname\relax
\typeout{** WARNING: IEEEtran.bst: No hyphenation pattern has been}%
\typeout{** loaded for the language `#1'. Using the pattern for}%
\typeout{** the default language instead.}%
\else
\language=\csname l@#1\endcsname
\fi
#2}}
\providecommand{\BIBdecl}{\relax}
\BIBdecl

\bibitem{mandianttrends2024}
\BIBentryALTinterwordspacing
Mandiant, ``Special report: Mandiant m-trends 2024,'' 2024. [Online]. Available: \url{https://services.google.com/fh/files/misc/m-trends-2024.pdf}
\BIBentrySTDinterwordspacing

\bibitem{ladisaDetectionMaliciousJava2022}
P.~Ladisa, H.~Plate, M.~Martinez, O.~Barais, and S.~E. Ponta, ``Towards the {{Detection}} of {{Malicious Java Packages}},'' in \emph{Proceedings of the 2022 {{ACM Workshop}} on {{Software Supply Chain Offensive Research}} and {{Ecosystem Defenses}}}.\hskip 1em plus 0.5em minus 0.4em\relax {Los Angeles CA USA}: {ACM}, 2022, pp. 63--72.

\bibitem{10.1145/3691621.3694950}
C.~Liang, Q.~Wei, Z.~Jiang, Y.~Wang, and J.~Du, ``A source code vulnerability detection method based on adaptive graph neural networks,'' in \emph{Proceedings of the 39th IEEE/ACM International Conference on Automated Software Engineering Workshops}, ser. ASEW '24.\hskip 1em plus 0.5em minus 0.4em\relax New York, NY, USA: Association for Computing Machinery, 2024, p. 187–196.

\bibitem{OpenSourceDatasetMaliciousSoftwarePackages}
\BIBentryALTinterwordspacing
D.~S. Labs, ``malicious-software-packages-dataset,'' Mar 2023. [Online]. Available: \url{https://github.com/datadog/malicious-software-packages-dataset}
\BIBentrySTDinterwordspacing

\bibitem{10.1007/978-3-030-52683-2_2}
M.~Ohm, H.~Plate, A.~Sykosch, and M.~Meier, ``Backstabber’s knife collection: A review of open source software supply chain attacks,'' in \emph{Detection of Intrusions and Malware, and Vulnerability Assessment: 17th International Conference, DIMVA 2020, Lisbon, Portugal, June 24–26, 2020, Proceedings}.\hskip 1em plus 0.5em minus 0.4em\relax Berlin, Heidelberg: Springer-Verlag, 2020, p. 23–43.

\bibitem{trandetectvul2024}
H.-C. Tran, A.-D. Tran, and K.-H. Le, ``{DetectVul}: {A} statement-level code vulnerability detection for {Python},'' \emph{Future Generation Computer Systems}, p. 107504, Sep. 2024.

\bibitem{malicious-packages}
O.~S.~S. Foundation, ``Openssf malicious packages,'' \url{https://github.com/ossf/malicious-packages/}, 2024.

\bibitem{tan_2024_14197378}
\BIBentryALTinterwordspacing
Z.~Tan, C.~Anagnostopoulos, and J.~Singer, ``{OSPtrack: A Labelled Dataset Targeting Simulated Open-Source Package Execution},'' Nov. 2024. [Online]. Available: \url{https://doi.org/10.5281/zenodo.14197378}
\BIBentrySTDinterwordspacing

\bibitem{9382367}
S.~Peisert, B.~Schneier, H.~Okhravi, F.~Massacci, T.~Benzel, C.~Landwehr, M.~Mannan, J.~Mirkovic, A.~Prakash, and J.~B. Michael, ``Perspectives on the solarwinds incident,'' \emph{IEEE Security \& Privacy}, vol.~19, no.~2, pp. 7--13, 2021.

\bibitem{ohmBackstabberKnifeCollection2020}
M.~Ohm, H.~Plate, A.~Sykosch, and M.~Meier, ``Backstabber's {{Knife Collection}}: {{A Review}} of {{Open Source Software Supply Chain Attacks}},'' 2020.

\bibitem{zhang_malicious_2023}
J.~Zhang, K.~Huang, B.~Chen, C.~Wang, Z.~Tian, and X.~Peng, ``\BIBforeignlanguage{en}{Malicious {Package} {Detection} in {NPM} and {PyPI} using a {Single} {Model} of {Malicious} {Behavior} {Sequence}},'' Sep. 2023, arXiv:2309.02637 [cs].

\bibitem{guo2023empiricalstudymaliciouscode}
\BIBentryALTinterwordspacing
W.~Guo, Z.~Xu, C.~Liu, C.~Huang, Y.~Fang, and Y.~Liu, ``An empirical study of malicious code in pypi ecosystem,'' 2023. [Online]. Available: \url{https://arxiv.org/abs/2309.11021}
\BIBentrySTDinterwordspacing

\bibitem{rozi_securing_2024}
M.~F. Rozi, T.~Ban, S.~Ozawa, A.~Yamada, T.~Takahashi, and D.~Inoue, ``\BIBforeignlanguage{en}{Securing {Code} with {Context}: {Enhancing} {Vulnerability} {Detection} through {Contextualized} {Graph} {Representations}},'' \emph{\BIBforeignlanguage{en}{IEEE Access}}, pp. 1--1, 2024.

\bibitem{tran_detectvul_2024}
H.-C. Tran, A.-D. Tran, and K.-H. Le, ``{DetectVul}: {A} statement-level code vulnerability detection for {Python},'' \emph{Future Generation Computer Systems}, p. 107504, Sep. 2024.

\bibitem{Parisot2024RansomwareDL}
A.~Parisot, L.~M.~S. Bento, and R.~C.~S. Machado, ``Ransomware detection: Leveraging sandbox, text mining techiques and machine learning,'' \emph{2024 IEEE International Workshop on Metrology for Industry 4.0 \& IoT (MetroInd4.0 \& IoT)}, pp. 446--451, 2024.

\bibitem{juan_a__herrera_silva__2023}
J.~A.~H. Silva and M.~H. Álvarez, ``Dynamic feature dataset for ransomware detection using machine learning algorithms,'' \emph{Sensors}, vol.~23, no.~3, pp. 1053--1053, 2023.

\bibitem{landauer_have_2021}
M.~Landauer, F.~Skopik, M.~Wurzenberger, W.~Hotwagner, and A.~Rauber, ``\BIBforeignlanguage{en}{Have it {Your} {Way}: {Generating} {Customized} {Log} {Datasets} {With} a {Model}-{Driven} {Simulation} {Testbed}},'' \emph{\BIBforeignlanguage{en}{IEEE Transactions on Reliability}}, vol.~70, no.~1, pp. 402--415, Mar. 2021.

\bibitem{aizawa2003information}
A.~Aizawa, ``An information-theoretic perspective of tf--idf measures,'' \emph{Information Processing \& Management}, vol.~39, no.~1, pp. 45--65, 2003.

\bibitem{10.1145/2976749.2989064}
B.~Sun, A.~Fujino, and T.~Mori, ``Poster: Toward automating the generation of malware analysis reports using the sandbox logs,'' in \emph{Proceedings of the 2016 ACM SIGSAC Conference on Computer and Communications Security}, ser. CCS '16.\hskip 1em plus 0.5em minus 0.4em\relax New York, NY, USA: Association for Computing Machinery, 2016, p. 1814–1816.

\bibitem{10529261}
R.-V. Mahmoud, M.~Anagnostopoulos, S.~Pastrana, and J.~M. Pedersen, ``Redefining malware sandboxing: Enhancing analysis through sysmon and elk integration,'' \emph{IEEE Access}, vol.~12, pp. 68\,624--68\,636, 2024.

\bibitem{shuofei_zhu__2020}
S.~Zhu, Z.~Zhang, L.~Yang, L.~Song, and G.~Wang, ``Benchmarking label dynamics of virustotal engines.''\hskip 1em plus 0.5em minus 0.4em\relax Association for Computing Machinery, 2020, pp. 2081--2083.

\bibitem{barr-smithExorcistAutomatedDifferential2022}
F.~{Barr-Smith}, T.~Blazytko, R.~Baker, and I.~Martinovic, ``Exorcist: {{Automated Differential Analysis}} to {{Detect Compromises}} in {{Closed-Source Software Supply Chains}},'' in \emph{Proceedings of the 2022 {{ACM Workshop}} on {{Software Supply Chain Offensive Research}} and {{Ecosystem Defenses}}}.\hskip 1em plus 0.5em minus 0.4em\relax {Los Angeles CA USA}: {ACM}, 2022, pp. 51--61.

\bibitem{TANG2022108261}
B.~Tang, J.~Wang, Z.~Yu, B.~Chen, W.~Ge, J.~Yu, and T.~Lu, ``Advanced persistent threat intelligent profiling technique: A survey,'' \emph{Computers and Electrical Engineering}, vol. 103, p. 108261, 2022.

\bibitem{9834133}
Y.~Ren, Y.~Xiao, Y.~Zhou, Z.~Zhang, and Z.~Tian, ``Cskg4apt: A cybersecurity knowledge graph for advanced persistent threat organization attribution,'' \emph{IEEE Transactions on Knowledge and Data Engineering}, vol.~35, no.~6, pp. 5695--5709, 2023.

\bibitem{LIU2024102748}
R.~Liu, Y.~Wang, H.~Xu, J.~Sun, F.~Zhang, P.~Li, and Z.~Guo, ``Vul-lmgnns: Fusing language models and online-distilled graph neural networks for code vulnerability detection,'' \emph{Information Fusion}, p. 102748, 2024.

\end{thebibliography}
\end{document}